\documentclass[aps,prl,showpacs,twocolumn]{revtex4}
\usepackage{amsmath,amssymb,graphicx,epsfig,color,bbm}

\def\nn{\nonumber}
\def\llangle{\langle\!\langle}
\def\rrangle{\rangle\!\rangle}

\begin{document}

\title{Robust Distant Entanglement Generation Using Coherent Multiphoton Scattering}

\author{Ching-Kit Chan and L. J. Sham}
\affiliation{Department of Physics, Center for Advanced Nanoscience, University of California San Diego, La Jolla, California 92093-0319, USA}
\date{\today}

\begin{abstract}
We describe a protocol to entangle two qubits at a distance by using resonance fluorescence. The scheme makes use of the postselection of large and distinguishable fluorescence signals corresponding to entangled and unentangled qubit states, and has the merits of both high success probability and high entanglement fidelity owing to the multiphoton nature. Our result shows that the entanglement generation is robust against photon fluctuations in the fluorescence signals for a wide range of driving fields. We also demonstrate that this new protocol has an average entanglement duration within the decoherence time of corresponding qubit systems, based on current experimental photon efficiency.
\end{abstract}


\maketitle


The generation and controllability of entanglement between distant quantum states have been at the heart of quantum computation and quantum information processing. Since the early proposal of entanglement generation in atomic ensemble systems \cite{duan01}, many theoretical ideas have been put forward \cite{childress05,duan06,waks09,duan10} and experimental demonstrations of distant entanglement have been recently performed in trapped ion systems \cite{moehring07,maunz09,olmschenk09}, with an average entanglement fidelity up to $90\%$. These existing approaches to generate entanglement are based on postselection using single-photon measurement \cite{duan01,childress05,duan06} and have the merit of high fidelity entanglement creation. However, due to the single-photon inefficiency, these protocols have a rather low success probability in practice ($\sim 10^{-8}$  in trapped ion experiments \cite{duan10}). This limitation leads to a very long average entanglement time compared to the decoherence time of the qubit system. To counter this difficulty, much experimental effort has been devoted to improving the single-photon efficiency in various qubit systems \cite{shu10,englund10,sage12}.

Alternative theoretical proposals to generate entanglement make use of Raman transitions of qubit systems embedded in a cavity \cite{vanloock06,vanloock08}. This scheme utilizes bright coherent light and thus is expected to be more efficient than the single-photon protocols, but at the expense of a moderate entanglement fidelity due to the cavity loss and spontaneous emission. Some solutions to overcome the sensitivity to spontaneous emission have been suggested replacing the coherent input light by either the Fock or the NOON states \cite{huang08}, which is experimentally more demanding. Therefore, a dichotomy exists in the architecture of entanglement generation that is either based on a low rate, high fidelity single-photon measurement, or a more effective, coherent photon protocol that has comparatively less fidelity.

To tackle such a dilemma, in this Letter we introduce a robust distant entanglement protocol using resonance fluorescence. By coupling the laser photons with the qubits in a Mach-Zehnder-type interferometer, the outgoing fluorescence signals can have a very different number of photons depending on the entanglement status of the qubits. We can thereby achieve entanglement between two distant qubits through the postselection of the detected many-photon states. This multiphoton method does not require any single-photon detector or sensitive phase measurement, and thus it has a higher rate of success without sacrificing the entanglement fidelity. Fluorescence is highly effective for its detection \cite{myerson08} and for this reason, has been used in the proposal of entanglement distribution in quantum repeaters once entanglement is created \cite{brask10}. Our scheme takes the same advantage of fluorescence and only involves the detection and distinguishabilty of the many-photon fluorescence signals, feasible using current experimental technology.

\begin{figure}[t]
\begin{center}
\includegraphics[angle=0, width=1.0\columnwidth]{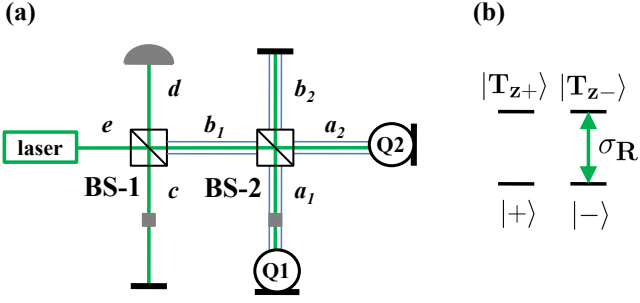}
\caption{(color online) (a) Schematic for the entanglement of distant qubits based on multiphoton scattering. The qubits are confined in a mirror-ended waveguide system. The grey boxes in $a_1$ and $c$ modes are $\pi/2$ phase shifters. The qubits are driven by a circularly
polarized laser at resonance and the scattered fluorescence signals interfere at BS-2. Due to the optical selection rule, the fluorescence signal serves as a fingerprint of the state of the electronic spin such that its detection at mode $d$ would postselectively project the two qubits state to an entangled state. (b) provides an example of energy levels in the QD qubit system, where the laser only selectively couples the $-3/2$ trion state $\left| T_{z-} \right \rangle$ and the $-1/2$ spin state $\left| - \right \rangle$.}
\label{fig_setup}
\end{center}
\end{figure}

Figure~\ref{fig_setup}(a) shows our setup to entangle two distant qubits. For illustrative purpose, we consider the singly charged quantum dot (QD) as our qubit system \cite{xu07,xu08}, where each dot consists of four energy levels as shown in Fig.~\ref{fig_setup}(b). Direct generalization to other qubit systems, such as the diamond nitrogen-vacancy center, is straightforward. The two electronic spin states $\left| \pm \right \rangle$ form the qubit states, while only the $\left| - \right \rangle$ state would be optically coupled to the spin $-3/2$ trion state $\left| T_{z-} \right \rangle$ through the right-handed circularly polarized laser due to the selection rules. Our scheme consists of two main steps: (1) to postselect the qubit states by photon measurement at $t=\tau$, and (2) to recover the entangled state coherence for $t\geq 2\tau$. In the first step, the laser is split by the polarization-independent $50/50$ beam splitters (BS) \cite{ma11} to drive the two distant qubits embedded in the mirror-ended waveguide or optical fiber system. Owing to the state dependence of the resonance fluorescence, the scattered multiphoton state is entangled with the qubit system. Considering that each qubit is initially prepared in the state $\left| x+ \right \rangle = \left( \left| + \right \rangle + \left| - \right \rangle \right) /\sqrt 2$, the state of the qubits and the outgoing $a$ mode photons is given by
\begin{eqnarray}
\frac{1}{2}  \left[\begin{array}{l}
\begin{array}{l}
\left| ++ \right \rangle \left| i\alpha;- \alpha \right \rangle + \left| -- \right \rangle \left| i f_{\alpha}; - f_{\alpha} \right \rangle \\
\ \ \  + \left| +- \right \rangle \left| i\alpha;-f_{\alpha} \right \rangle
+ \left| -+ \right \rangle \left| i f_{\alpha};- \alpha \right \rangle \end{array}\end{array}\right],
\label{eq_wavefunction}
\end{eqnarray}
where each term gives the spin state of qubit 1, 2 and photon state in $a_1$, $a_2$ mode, $\left| \alpha \right\rangle$ is the coherent Glauber state that describes the laser, and $\left| f_{\alpha} \right\rangle = \left\langle - |e^{-iHt}  |-\rangle |\alpha \right\rangle$ denotes both the laser and fluorescence photon state governed by the electric-dipole Hamiltonian $H$. Note that $\left| f_{\alpha} \right\rangle$ is not a Glauber state and its BS transformation is not just a simple photon amplitude transformation. Then, the $a_1$ and $a_2$ modes, consisting of laser photons and fluorescence signals, are transformed to $b_1$ and $b_2$ modes by BS-2. The $b_1$ mode is further combined with the reflected laser mode $c$ at BS-1 such that the detector mode $d$ only contains fluorescence photons (see the Supplemental Material \cite{supple}).

Postselection of the qubit state can be achieved by measuring the number of photons in the detector mode. We shall see below that, due to the BS, the multiphoton states associated with $\left| +- \right \rangle$ and $\left| -+ \right \rangle $ share the same number of photons in $b_1$ and thus $d$ modes, which are macroscopically different from those of the other two states. This allows us to project the qubit state on $\left| +- \right \rangle$ and $\left| -+ \right \rangle$ by photon measurement at $d$. At this point, these two states have the same photon state in the $b_1$ mode, but have a $\pi$ phase difference in the $b_2$ mode. Therefore, in the second step, similar to the spin echo technique, we apply a $\pi$ pulse to flip both spins at $t=\tau$ and continue the fluorescence process until $t=2\tau$, at which time the qubit states would share the same photon state. By tracing out the photon degrees of freedom,  we achieve the maximally entangled state $\left | E\right\rangle = \left( \left| +- \right \rangle + \left| -+ \right \rangle \right) / \sqrt 2$ for $t \geq 2 \tau$. This postselection scheme shares the same high entanglement fidelity as in the single-photon protocol, but exponentially boosts the entanglement efficiency as only multiphoton measurement is involved.

We will focus on the postselection process in the following. The many-photon resonance fluorescence establishes a robust entanglement between the photons and the qubits. The photon dynamics in the $a$ modes is governed by
\begin{eqnarray}
a_{s,k}(t)=e^{-i\omega_k t} a_{s,k}(0) -i g_k \int_0^t dt' \sigma_{s,-} (t') e^{-i\omega_k  (t-t')},
\label{eq_photon}
\end{eqnarray}
where $\omega_k$ and $g_k$ are, respectively, the photon frequency and dipole-electric coupling, $\sigma_- = \left| - \right \rangle \left\langle T_{z-} \right |$, and $s=1,2$. The first and second terms correspond to the laser and fluorescence photons, respectively. For our process, the entanglement is heralded by the photons arriving at the detector $d$, different from the single-photon scenario that requires a simultaneous photon registration by two detectors.

\begin{figure}[t]
\begin{center}
\includegraphics[angle=0, width=1.0\columnwidth]{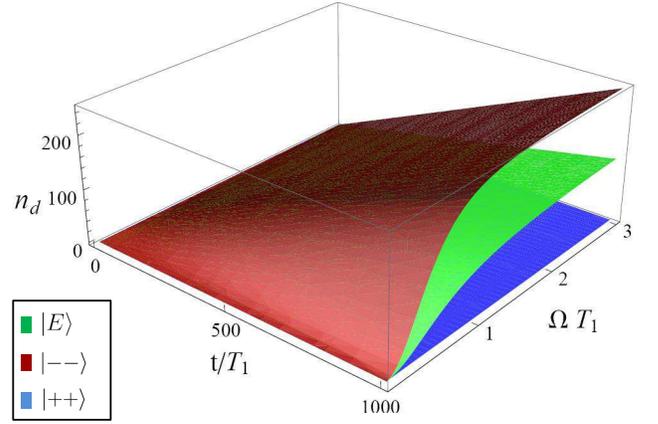}
\caption{(color online) Time and Rabi frequency dependences of the average number of photon detected at $d$ for $\eta=1$. The state-selective fluorescence signals increase linearly with time when $t \gg T_1$ and are distinct for state postselection. }
\label{fig_photon}
\end{center}
\end{figure}

Based on the BS transformations, the state-selective average number of fluorescence photons being detected at $d$ can be evaluated by solving the optical Bloch equation \cite{scully97}. The results in the long time regime ($t \gg T_1$) can be analytically expressed as
\begin{eqnarray}
\label{eq_photon number}
n_{d;++} (t)&=& 0, \nn \\
n_{d;E}(t) &=& \frac{\eta }{4} \frac{\left(\Omega T_1\right)^2}{1+2\left(\Omega T_1\right)^2} \frac{t}{T_1},  \\
n_{d;--} (t)&=& \eta \frac{\left(\Omega T_1\right)^2+\left(\Omega T_1\right)^4}{\left(1+2\left(\Omega T_1\right)^2\right)^2} \frac{t}{T_1},\nn
\end{eqnarray}
where $T_1$ is the relaxation time for the transition $\left|-\right\rangle \leftrightarrow \left|T_{z-}\right\rangle$, $\Omega$ is the Rabi frequency, and $\eta$ is the photon collection and detection efficiency. Figure~\ref{fig_photon} plots the maximum number of detected photons (i.e., $\eta=1$) as a function of time and Rabi frequency of the qubit. Large numbers of photons are generated conditioned on the qubit states and increase linearly with time when $t\gg T_1$. The distinguishability between $n_{d;E}$ and the other two numbers of photons in most of the region allows a probabilistic entanglement generation by postselection with an ideal probability of $1/2$, compared to $1/4$ in the single-photon protocol. We note that the dark transitions due to the hole mixing in QD system \cite{liu10}, which would lead to an estimated $1\%$ of fluorescence photons, are negligible in this many-photon entanglement scheme.

\begin{figure}[t]
\begin{center}
\includegraphics[angle=0, width=1.0\columnwidth]{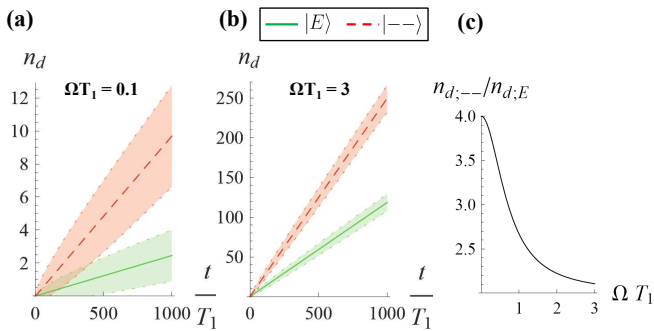}
\caption{(color online) (a) and (b) provide the fluctuations in the number of detected fluorescence photons for various field strengths. Shaded areas refer to the region of $n_{d} \pm \Delta n_{d}$. Signals corresponding to entangled and unentangled states can be easily separated with a high fidelity in both the low and high field domains. (c) gives the ratio of $n_{d;--}/n_{d;E}$ as a function of the Rabi frequency for $t \gg T_1$.}
\label{fig_noise}
\end{center}
\end{figure}

We now analyze the noise of our entanglement scheme. Unlike the single-photon protocol whose fidelity is limited by dark counts in the experiment, our multiphoton entanglement approach is not sensitive to the background noise, but relies on a clear distinction between large fluorescence signals. Photon statistics in resonance fluorescence has been well studied since Mandel \cite{mandel79} and has been verified in atomic experiments \cite{short83}. We make use of the same approach by Mandel in our entanglement process and compute the fluctuations in the number of detected photons by \cite{mandel79}
\begin{eqnarray}
\Delta n_{d} (t)^2&=& n_{d} (t) + \frac{\eta^2 }{8 T_1^2}\sum_{s_1,s_2,s_3,s_4} \int_0^t dt_2\int_0^{t_2} dt_1  \nn \\
&\times&\left\llangle  \sigma_{s_1,+}(t_1) \sigma_{s_2,+}(t_2)\sigma_{s_3,-}(t_2) \sigma_{s_4,-}(t_1)\right\rrangle,
\end{eqnarray}
where $\left\llangle \sigma_1 \sigma_2 \sigma_3 \sigma_4\right\rrangle = \left\langle \sigma_1 \sigma_2 \sigma_3 \sigma_4 \right\rangle-\left\langle \sigma_1 \sigma_4 \right\rangle \left\langle \sigma_2 \sigma_3\right\rangle$ and we have made the phase change $\pm i \sigma_{2,\pm} \rightarrow \sigma_{2,\pm}$.
The four-point correlation functions can be evaluated using the quantum regression theorem \cite{scully97}, being applicable for our qubit system. Note that this expression has already taken into account the noise of the nonideal detector.

\begin{figure}[t]
\begin{center}
\includegraphics[angle=0, width=1.0\columnwidth]{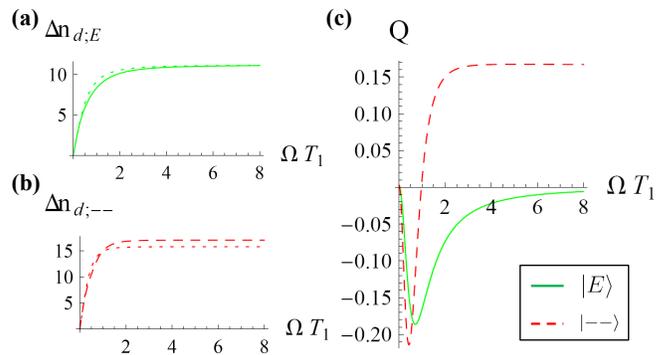}
\caption{(color online) (a) and (b) show the field dependence of photon fluctuations for entangled and unentangled states at $t/T_1=10^3$. Dotted lines represent Poissonian fluctuations. $\Delta n_{d;E}$ is entirely sub-Poissonian, while $\Delta n_{d;--}$ can be either sub-Poissonian or super-Poissonian depending on $\Omega T_1$. (c) provides the corresponding Mandel parameters. The super-Poissonian statistics is not significant enough to destroy the distinguishability of fluorescence signals.}
\label{fig_noisecomparison}
\end{center}
\end{figure}

Figure~\ref{fig_noise}(a) and~\ref{fig_noise}(b) illustrate how the noise of the maximally detected fluorescence photons depends on the qubit state. It is clear that the many-photon fluorescence signals are distinguishable in both low field (Rayleigh) and high field (Mollow) regimes, even in the presence of photon number fluctuation. Figure~\ref{fig_noise}(c) shows the driving field dependence of the ratio of $n_{d;--}$ to $n_{d;E}$. Even though this ratio is larger when the field is low, a high field is preferred for a better signal to noise ratio. A long time operation is also necessary for the same reason. The fluorescence signal saturates at $\Omega T_1 \sim 3$ and approaches to the asymptotic values $n_{d;E}(t) \rightarrow t/(8 T_1)$ and $n_{d;--}(t) \rightarrow t/(4 T_1)$ at a larger field strength.

Figure~\ref{fig_noisecomparison} details the field dependence of the noise statistics of photons for the entangled and unentangled states for $\eta=1$. At large time ($t/T_1=10^3$), the photon signal corresponding to the entangled state demonstrates a sub-Poissonian statistics as shown in Fig.~\ref{fig_noisecomparison}(a). This result is just the same as that of the standard resonance fluorescence, because the entangled state only consists of fluorescence photons coming from either one of the qubits. By contrast, the photons associated with the unentangled state $\left |-- \right\rangle$ exhibit a transition from sub-Poissonian to super-Poissonian behavior when the field increases. Figure~\ref{fig_noisecomparison}(c) shows that the Mandel parameter, $Q=(\Delta n_{d})^2/n_{d}-1$, for the unentangled state, can be as low as that of the entangled state in the low field regime and becomes positive to the asymptotic value of $1/6$ in the large field limit. Note that all these non-Poissonian results are in agreement with the non-Glauber character of the fluorescence state mentioned above. We remark that the Mandel parameter is limited by the photon efficiency, so that $|Q| \sim \eta \ll 1$ and, therefore, the measured photon statistics will be very close to Poissonian in the realistic experimental situation.

An important aspect of all the entanglement mechanism is the average entanglement rate. The low efficiency of the entanglement protocol using single-photon detection severely limits the rate of each successful operation. For instance, in the existing trapped ions experiment \cite{duan10}, the success probability is $\sim 2.2 \times 10^{-8}$, leading to an average entanglement time around $600~\text{s}$, being longer than the decoherence time of the trapped ion qubit. This posts a dramatic restriction to the scalability of the corresponding qubit systems. While experimental technique is advancing to increase the single-photon efficiency, our entanglement generation scheme based on the many-photon resonance fluorescence signal can help to solve the problem and can have a success probability close to unity. We discuss our entanglement rate in the following. Our entanglement scheme depends on the distinction of fluorescence signals corresponding to entangled and unentangled states (Fig.~\ref{fig_noise}). In the large field region ($\Omega T_1 > 3$), these two signals are already well separated with a confidence level (or fidelity) being more than $90\%$, when $\eta \times t / T_1 \approx 130$. In other words, we can reach the ideal success probability (i.e., $1/2$) using an operational time that is long enough to produce distinguishable multiphoton states and entangle the qubits, but short compared with the spin decoherence time. Taking the trapped ion example, the photon collection and detection efficiencies are $2\%$ and $15\%$, respectively, so that $\eta=3\times 10^{-3}$. This corresponds to our average entanglement time being $\sim 1.7 \times 10^5~T_1 \approx 1.4~\text{ms}$ ($T_1$ being the spontaneous emission time for the $^2 P_{1/2}$ to $^2 S_{1/2}$ transition in the trapped ion qubit), which is five orders less than that of the single-photon protocol and more importantly, much shorter than the decoherence time of the trapped ion system. Similarly, for the nitrogen-vacancy center qubit system with a similar spontaneous emission time \cite{togan10}, an average entanglement time of milliseconds can be achieved under the same photon efficiency and is comparable to its coherence time scale.

The same is true for the QD qubits in spite of the relatively shorter decoherence time. In the singly charged QD system, the upper trion state roughly has the relaxation time $T_1 \approx 0.1~\text{ns}$ \cite{xu07}. Taking realistic collection ($6.7\%$) and detection ($15\%$) efficiencies in the resonance fluorescence experiments \cite{short83}, our multiphoton entanglement scheme has an average entanglement time $\sim 5.2~\text{$\mu$s}$, being comparable to the electron coherence time achieved in the spin echo experiment of a single QD \cite{press10}. Such a coherence time scale can also be reached by using the coherent dark-state spectroscopy \cite{xu09,sun12}. This many-photon entanglement scheme has to be contrasted with the single-photon protocol that would instead result in an estimated time to entangle two distant QDs to be at least of the order of milliseconds on average, based on existing physical parameters. In other words, unlike the single-photon entanglement protocol that requires a sharp improvement in the experimental performance of single-photon measurements, the multiphoton approach is more suitable for a scalable qubit entanglement under realistic experimental conditions.

Another challenge to the distant qubit entanglement by postselection is the mismatch of the optical characters of the two qubits. Similar to the single-photon protocol, our scheme also requires the indistinguishability of each photon from both qubits, i.e., the discrepancy of their optical frequencies being less than the relaxation rate. On the other hand, consider two qubit systems having different resonant Rabi frequencies ($\Omega, \Omega'$) and relaxation times ($T_1,T_1'$). According to Eq.~(\ref{eq_photon number}), the large field behaviors of the fluorescence signals in this situation are that $n_{d;++}=0$, $n_{d;+-} \approx \eta~t /8T_1 $, $n_{d;-+} \approx \eta~t/8T_1' $, and $n_{d;--} \approx (\eta~t/8) (1/T_1 +1/T_1')$. This means that our entanglement method is insensitive to the Rabi frequency mismatch, and the $\left|-- \right\rangle$ state can still be differentiated from the other states for a moderate difference in $T_1$. In order to maintain the coherence of the entangled state, the difference in the corresponding fluorescence signals cannot be too large, i.e., $\left| n_{d;+-}-n_{d;-+}\right| < \Delta n_{d;+-(-+)}$. For $\eta t/ T_1\approx 130$, this roughly corresponds to $\sim 10\% $ relative discrepancy in the relaxation rate. In this regime, a high rate and fidelity entanglement is still feasible.

In summary, we have introduced a new entanglement generation scheme using postselection of coherent multiphoton signals. It improves on the single- photon protocols by increasing the success probability and the average entanglement rate considerably, while maintaining the high fidelity of entanglement. Without significant enhancement in single-photon measurement efficiency, this proposed multiphoton protocol already allows an average entanglement of two distant qubits before their decoherence based on current experimental technologies, while experimental improvements would further reduce the operational duration. We believe this would open a new direction in the design of hybrid multiphoton-qubit network in quantum information processing.

\begin{acknowledgements}
This research was supported by the U.S. Army Research Office MURI Grant No. W911NF0910406 and by NSF Grant No. PHY-1104446. C.K.C. thanks Guy Cohen and Bo Sun for useful discussions.
\end{acknowledgements}


\end{document}


\title{Supplemental Material for ``Robust distant-entanglement generation using coherent multiphoton scattering"}

\author{Ching-Kit Chan and L. J. Sham}
\affiliation{Department of Physics, Center for Advanced Nanoscience, University of California San Diego, La Jolla, California 92093-0319, USA}
\date{\today}

\maketitle


The qubit-photon state evolution during the entanglement procedure is detailed in the following. Referring to Fig.~1(a) of the paper, we define the laser coherent state at mode $a$ as $\left| D_a(\alpha)\right\rangle = \hat D_a(\alpha)\left| 0\right\rangle$, where $\hat D$ is the displacement operator; and the fluorescence state including the laser photon at mode $a'$ as $\left| F_{a'}(f_\alpha)\right\rangle = \left\langle-\right|e^{-iHt} \left|-\right\rangle\left| D_{a'}(\alpha)\right\rangle$, such that
\begin{eqnarray}
\left\langle D_a(\alpha)\right|a \left|D_a(\alpha)\right\rangle &=& \alpha, \nn \\ 
\left\langle F_{a'}(f_\alpha)\right|a' \left|F_{a'}(f_\alpha)\right\rangle &=& \alpha + f_\alpha,
\label{eq_1}
\end{eqnarray}
and the property that $\left| F_{a'}(e^{i\theta} f_\alpha)\right\rangle = \left| F_{a'}( f_{\alpha e^{i\theta}})\right\rangle$.

To begin with, the BSs split the laser into products of ingoing  coherent photon states, so that the initial state describing the qubit and the outgoing photon modes ($a_1$, $a_2$, $c$) is $\frac{1}{2}  (\left|++\right\rangle + \left|+-\right\rangle+ \left|-+\right\rangle+ \left|--\right\rangle)\times \left|D_{a_1}(i\alpha) D_{a_2} (-\alpha) D_{c}(i\sqrt 2\alpha)\right\rangle$. At time $0$, each qubit starts to fluoresce and entangles with the multiphoton state owing to the optical selection rule. The state just before BS-2 is given by:
\begin{eqnarray}
\frac{1}{2}  \left[\begin{array}{l}
\begin{array}{l}
\ \ \left| ++ \right \rangle \left| D_{a_1}(i\alpha) D_{a_2}(- \alpha) \right \rangle \\
+ \left| -- \right \rangle \left| F_{a_1}(i f_{\alpha})  F_{a_2}(- f_{\alpha})\right \rangle \\
+ \left| +- \right \rangle \left| D_{a_1}(i\alpha) F_{a_2}(- f_{\alpha}) \right \rangle \\
+ \left| -+ \right \rangle \left| F_{a_1}(i f_{\alpha})  D_{a_2}(- \alpha)\right \rangle 
\end{array}\end{array}\right],
\label{eq_2}
\end{eqnarray}
where the common coherent photon state at mode $c$ is not shown in this equation. Using the abbreviations $\left| \alpha \right\rangle = \left| D(\alpha)\right\rangle$ and $\left| f_\alpha \right\rangle = \left| F(f_\alpha)\right\rangle$, we recover Eq.~(1) of the paper. Next, the BS-2 transforms the $a_1$, $a_2$ modes to the $b_1$, $b_2$ modes. The state just after BS-2 now becomes:
\begin{eqnarray}
\frac{1}{2}  \left[\begin{array}{l}
\begin{array}{l}
\ \ \left| ++ \right \rangle \left| D_{b_1}(-\sqrt 2 \alpha) D_{b_2}(0) \right \rangle \\
+ \left| -- \right \rangle \left| F_{\frac{b_1+i b_2}{\sqrt 2}}(-f_{\alpha})  F_{\frac{b_1-i b_2}{\sqrt 2}}(- f_{\alpha})\right \rangle \\
+ \left| +- \right \rangle \left| D_{\frac{b_1+i b_2}{\sqrt 2}}(-\alpha)  F_{\frac{b_1-i b_2}{\sqrt 2}}(- f_{\alpha})\right \rangle \\
+ \left| -+ \right \rangle \left| F_{\frac{b_1+i b_2}{\sqrt 2}}(-f_{\alpha})  D_{\frac{b_1-i b_2}{\sqrt 2}}(-\alpha)\right \rangle \\
\end{array}\end{array}\right].
\label{eq_3}
\end{eqnarray}
Note that unlike the coherent Glauber state, the fluorescence state is not separable, i.e. $|F_{(a+a')/\sqrt 2}( f_{\alpha})\rangle \neq |F_{a}( f_{\alpha}/\sqrt2) F_{a'}( f_{\alpha}/\sqrt 2)\rangle$. After that, the $b_1$ mode is combined with the reflected laser mode $c$ at BS-1, so that the total state governing the qubits and all the output photons is:
\begin{widetext}
\begin{eqnarray}
\frac{1}{2}  \left[\begin{array}{l}
\begin{array}{l}
\ \ \left| ++ \right \rangle \left| D_{d}(0) D_{e}(-2\alpha) D_{b_2}(0) \right \rangle 
+ \left| -- \right \rangle \left| F_{- \frac{id}{2}+\frac{e}{2}+\frac{i b_2}{\sqrt 2}} (-f_\alpha)  F_{- \frac{id}{2}+\frac{e}{2}-\frac{i b_2}{\sqrt 2}} (-f_\alpha) D_{\frac{d-ie}{\sqrt 2}}(i\sqrt 2 \alpha) \right \rangle \\
+ \left| +- \right \rangle \left| D_{d}(\frac{i\alpha}{2}) D_{e}(-\frac{3\alpha}{2}) D_{b_2}(\frac{i\alpha}{\sqrt 2}) F_{- \frac{id}{2}+\frac{e}{2}-\frac{i b_2}{\sqrt 2}} (-f_\alpha) \right \rangle
+ \left| -+ \right \rangle \left| D_{d}(\frac{i\alpha}{2}) D_{e}(-\frac{3\alpha}{2}) D_{b_2}(\frac{-i\alpha}{\sqrt 2}) F_{- \frac{id}{2}+\frac{e}{2}+\frac{i b_2}{\sqrt 2}} (-f_\alpha) \right \rangle \\
\end{array}\end{array}\right].
\label{eq_4}
\end{eqnarray}
\end{widetext}

Based on the results of the paper, a postselection of the qubit state can be achieved by measuring the different number of fluorescence photons at mode $d$. Therefore, for our entanglement process, the state at time $\tau$ is projected to:
\begin{eqnarray}
\frac{1}{\sqrt 2}  \left[\begin{array}{l}
\begin{array}{l}
\ \ \left| +- \right \rangle \left| D_{\frac{b_1+i b_2}{\sqrt 2}}(-\alpha)  F_{\frac{b_1-i b_2}{\sqrt 2}}(- f_{\alpha})\right \rangle \\
+ \left| -+ \right \rangle \left| D_{\frac{b_1-i b_2}{\sqrt 2}}(-\alpha)  F_{\frac{b_1+i b_2}{\sqrt 2}}(-f_{\alpha}) \right \rangle \\
\end{array}\end{array}\right],
\label{eq_5}
\end{eqnarray}
where this equation is expressed in terms of the $b_1$ and $b_2$ photons for a simplified notation and the common $c$ mode is also not shown. It is clear that the states $\left| +- \right\rangle$ and $\left| -+ \right\rangle$ have the same $b_1$ photons, but a sign difference in the $b_2$ photons. Therefore, as described in the paper, we apply the spin echo technique at time $\tau$ and continue the fluorescence process until time $2\tau$, so that the roles of the additional photons in the states $\left| +- \right\rangle$ and $\left| -+ \right\rangle$ are reversed. For $t \geq 2 \tau$, by tracing out all the photon degrees of freedom, we achieve the pure and maximally entangled Bell state $\left |E\right\rangle = (\left| +- \right\rangle + \left| -+ \right\rangle )/\sqrt 2$.